\documentclass[12pt]{iopart}

\usepackage{amsmath,latexsym,amssymb,enumerate}
\usepackage{xcolor}
\usepackage[square,numbers]{natbib} 

\usepackage{graphicx}
\usepackage{natbib}
\usepackage{url}

\definecolor{dgreen}{rgb}{0,.5,0}
\definecolor{grau}{gray}{.5}
\definecolor{schwarz}{gray}{0}

\newcommand{\reff}[1]{(\ref{#1})}

\newcommand{\cg}[1]{\mathcal{#1}}

\newcommand{\av}[1]{\left|#1\right|}

\newcommand{\Ll}[1]{\left\|#1\right\|}
\newcommand{\N}[2]{\left\|#1\right\|_{#2}}

\newcommand{\brkts}[1]{\left(#1\right)}
\newcommand{\ebrkts}[1]{\left[#1\right]}

\newcommand{\brcs}[1]{\left\{#1\right\}}

\newcommand{\pd}[2]{\frac{\partial #1}{\partial #2}}

\newcommand{\bsplitl}[2]{
\begin{equation}
\begin{split}
#1
\end{split}
\label{#2}
\end{equation}}

\newcommand{\csection}[1]{%
  \section[#1]{\centering\Large\MakeUppercase\itshape #1}} 

\newtheorem{thm}{Theorem}[section]

\newtheorem{defn}[thm]{Definition}

\newtheorem{rem}[thm]{Remark}

\DeclareGraphicsExtensions{.png,.eps,.ps}

\begin{document}

\title[Upscaling phase field equations for periodic fluid flow]{Derivation of effective macroscopic Stokes-Cahn-Hilliard equations for periodic
immiscible flows in porous media}

\author{Markus Schmuck$^{1,2}$\footnote{Present address: School of Mathematical and Computer Sciences and the Maxwell Institute for Mathematical Sciences,
	Heriot-Watt University, EH14 4AS, Edinburgh, UK}, Marc Pradas$^{1}$, Grigorios A. Pavliotis$^2$, and Serafim Kalliadasis$^{1}$
}
	\address{
	$^1$ Department of Chemical Engineering,
	Imperial College London,
	South Kensington Campus,
	SW7 2AZ London, UK\\
	$^2$Department of Mathematics,
	Imperial College London,
	South Kensington Campus,
	SW7 2AZ London, UK
	}	
	\eads{\mailto{M.Schmuck@hw.ac.uk},\,\mailto{m.pradas@imperial.ac.uk},\,\mailto{g.pavliotis@imperial.ac.uk}, and \mailto{s.kalliadasis@imperial.ac.uk}}

\begin{abstract}
Using thermodynamic and variational principles we examine a basic phase field
model for a mixture of two incompressible fluids in strongly perforated
domains. With the help of the multiple scale method with drift and our
recently introduced splitting strategy for Ginzburg-Landau/Cahn-Hilliard-type
equations \cite{Schmuck2012a}, we rigorously derive an effective macroscopic phase field
formulation under the assumption of periodic flow and a sufficiently large
P\'eclet number. As for classical convection-diffusion problems, we obtain
systematically diffusion-dispersion relations (including
Taylor-Aris-dispersion).  Our results also provide a convenient analytical and computational
framework to macroscopically track interfaces in porous media. In view of the
well-known versatility of phase field models, our study proposes a promising
model for many engineering and scientific applications such as multiphase
flows in porous media, microfluidics, and fuel cells.
\end{abstract}

\submitto{\NL}
%

\maketitle

\csection{Introduction}\label{sec:Intr}
Fluid mixtures are ubiquitous in many scientific and engineering
applications. The dynamics of phase interfaces between fluids plays a central
role in rheology and hydrodynamics
\cite{Boyer2002,Chaikin1995,Feng2005,Liu2003,Probstein1994}. A recent attempt
of a systematic extension towards non-equilibrium two-phase systems is
\cite{Savin2012}, where the authors discuss the concept of local
thermodynamic equilibrium of a Gibbs interface in order to relax the global
thermodynamic equilibrium assumption. In \cite{Lowengrub1998}, it is shown
that the Cahn-Hilliard or diffuse interface formulation of a
quasi-compressible binary fluid mixture allows for topological changes of the
interface. Also of increasing interest is the mathematical and physical
understanding of wetting using diffuse interface
formulations~\cite{Sibley2013b,Sibley2013a} as well as wetting in the
presence of complexities such as electric fields~\cite{Lu2007,Eck2009}.

The study of flows in porous media is a delicate multiscale problem. This is
evident, for instance, by the fact, that the full problem without any
approximations is not computationally tractable with the present
computational power \cite{Adler1988,Jenny2003}. Also, from an empirical perspective the consideration of
the full multiscale problem is very challenging due to the difficulty of
obtaining detailed information about the pore geometry. These empirical and
computational restrictions strongly call for systematic and reliable
approximations which capture the essential physics and elementary dynamic
characteristics of the full problem in an averaged sense. A very common and
intuitive strategy is volume averaging \cite{Quintard1993,Whitaker1986}. The
method of moments \cite{Aris1956,Brenner1980} and multiple scale expansions
\cite{Bensoussans1978,Pavliotis2008,Cioranescu1999} have been used in this
context. The latter method is more systematic and reliable since it allows
for a rigorous mathematical verification. The volume averaging strategy still
lacks a consistent and generally accepted treatment of nonlinear terms.
Therefore, the multiscale expansion strategy is used as a basis for the
theoretical developments in the present study.

The celebrated works in~\cite{Aris1956,Brenner1980,Taylor1953} initiated an
increasing interest in the understanding of hydrodynamic dispersion on the
spreading of tracer particles transported by flow, with numerous
applications, from transport of contaminants in rivers to chromatogaphy. In
\cite{Rubinstein1986}, it is shown that the multiscale expansion strategy
allows to recover the dispersion relation found in \cite{Brenner1980}. The
study of multiphase flows in porous media is considerably more complex; see
e.g. the comprehensive review in~\cite{Adler1988} which still serves as a
basis for several studies in the field. The central idea for many approaches
to multiphase flows is to extend Darcy's law to multiple phases. With the
help of Marle's averaging method \cite{Marle1982} and a diffuse interface
model, effective two-phase flow equations are presented in
\cite{Papatzacos2002,Papa2010}. In \cite{Atkin1976}, Atkin and Craine even
present constitutive theories for a binary mixture of fluids and a porous
elastic solid. A combination of the homogenization method and a multiphase
extension of Darcy's law as a description of multiphase flows in porous media
is applied in the articles \cite{Bourgeat1996,Schweizer2008a}, for instance.

An application of increasing importance for a renewable energy infrastructure
are fuel cells~\cite{Promislow2009}. This article combines the complex
multiphase interactions with the help of the Cahn-Hilliard phase field method
and a total free energy characterizing the fuel cell. An upscaling of the
full thermodynamic model proposed in \cite{Promislow2009} is obviously very
involved due to complex interactions over different scales. In this context,
an upscaled macroscopic description of a simplified (i.e., no fluid flow and
periodic catalyst layer) is derived in \cite{Schmuck2012b,Schmuck2011}.

\medskip

Consider the total energy density for an
interface between two phases,
\bsplitl{
e({\bf x}({\bf X},t),t)
    :=
    \frac{1}{2}\av{
    	\frac{\partial {\bf x}({\bf X},t)}{\partial t}
    }^2
    -\frac{\lambda}{2}
    	\av{
		\nabla_{\bf x}\phi({\bf x}({\bf X},t),t))
	}^2
    -\frac{\lambda}{2}F(\phi({\bf x}({\bf X},t),t))
    \,,
}{FrEn} where $\phi$ is a conserved order-parameter that evolves between
different liquid phases represented as the minima of a homogeneous free
energy $F$. The parameter $\lambda$ represents the surface tension effect,
i.e. $\lambda\propto(\textrm{surface tension})\times(\textrm{capillary
width})=\sigma\eta$. 
The variable ${\bf X}$ stands
for the Lagrangian (initial) material coordinate and ${\bf x}({\bf X},t)$ represents the Eulerian
(reference) coordinate.
Our derivation (Section \ref{sec:Proof}) is valid for general
free energies $F$ and uses the method of an asymptotic multiscale expansion with drift \cite{Allaire2010a}.
Furthermore, we establish the wellposedness (Theorem \ref{thm:2}) of the
upscaled/homogenized equations for polynomial free energies of the
following form \cite{Temam1997}:

\medskip

{\bf Assumption (PF):} \emph{The free energy densities $F$ in \reff{FrEn} are polynomials
of order $2r-1$, i.e.,
\bsplitl{
f(u)
	= \sum_{i=1}^{2r-1}a_iu^i\,,
	\quad r\in\mathbb{N}\,,
	\quad r\geq 2
	\,,
}{PNfrEn}
with f(u)=F'(u) vanishing at $u=0$,
\bsplitl{
F(u)
	= \sum_{i=2}^{2r}b_iu^i\,,
	\quad ib_i=a_{i-1}\,,
	\quad 2\leq i\leq 2r
	\,,
}{f}
where the leading coefficient of both $F$ and $f$ is positive, i.e.,
$a_{2r-1}=2rb_{2r}>0$.
}

\medskip

\begin{rem}\label{rem:1} \emph{(Double-well potential)} Free energies $F$ satisfying the
\emph{Assumption (PF)} form a general class which also includes the double-well
potential for $r=2$ with $f(u)=-\alpha u+\beta u^3$, $\alpha,\,\beta>0$, for which \reff{PhMo}$_4$
is called the convective Cahn-Hilliard equation. We note that the double-well is scaled by
$\frac{1}{4\eta^2}$, i.e., $F(u)=\frac{1}{4\eta^2}(u^2-1)^2$ such that
one recovers the Hele-Shaw problem in the limit $\eta\to 0$ \cite{Hyon2010,Liu2003}.
\end{rem}

The last two terms in \reff{FrEn} form the well-known density of the
Cahn-Hilliard/Ginzburg-Landau phase field formulation adapted to the flow map ${\bf x}({\bf X},t)$ defined
by
\bsplitl{
\begin{cases}
\quad
\frac{\partial {\bf x}}{\partial t}
	= {\bf u}({\bf x}({\bf X},t),t)\,,
	& \textrm{}
\\\quad
{\bf x}({\bf X},0)
	= {\bf X}\,.
	&
\end{cases}
 }{FlMap}
The first term in \reff{FrEn} is the kinetic energy, which accounts for the fluid flow of incompressible materials, i.e.,
\bsplitl{
\begin{cases}
\quad
\frac{\partial {\bf u}}{\partial t}
	+({\bf u}\cdot\nabla){\bf u}
	-\mu\Delta{\bf u}
	+\nabla p
	= \pmb{\eta}\,,
	&
\\\quad
{\rm div}\,{\bf u}
	= 0\,,
	&
\end{cases}
}{incMat}
where we additionally added the second order term multiplied by the viscosity $\mu$. The variable $\pmb{\eta}$ is a driving
force acting on the fluid. We are interested in the mixture of two incompressible and immiscible fluids of the
same viscosity $\mu$. Hence, we can employ generic free energies \reff{FrEn}
showing a double-well form as is the case often in applications, e.g.~\cite{Pradas2012}.

Suppose that $\Omega\subset\mathbb{R}^d$, with $d>0$ the dimension of space, denotes the domain which is initially occupied by the fluid. Then, we can define for an arbitrary length of time
$T>0$ the total energy by
\bsplitl{
E({\bf x})
	:=\int_0^T\int_{\Omega}
		e({\bf x}({\bf X},t),t)
	\,d{\bf X}\,dt
	\,. }{Action} The energy \reff{Action} combines an action functional for
the flow map ${\bf x}({\bf X},t)$ and a free energy for the order parameter
$\phi$. This combination of mechanical and thermodynamic energies seems to go
back to \cite{deGennes1993,Doi1986,Lin1995,Liu2003,Lowengrub1998}.
Subsequently, we will focus on quasi-stationary, i.e., ${\bf u}_t={\bf 0}$
and $\pmb{\eta}\neq {\bf 0}$, and low-Reynolds number flows such that we
can neglect the nonlinear term $({\bf
u}\cdot\nabla){\bf u}$. Then, classical ideas from the calculus of
variations \cite{Struwe2008} and the theory of gradient flows together with
an imposed wetting boundary condition $\int_{\partial\Omega}g({\bf
x})\,do({\bf x})$ for $g({\bf x})\in H^{3/2}(\partial\Omega)$ lead to the
following set of equations
\bsplitl{
\textrm{(Homogeneous case)}\,\,
\begin{cases}
-\mu\Delta{\bf u}
	+\nabla p
	= \pmb{\eta}
	& \textrm{in }\Omega_T
	\,,
\\\quad
{\rm div}\,{\bf u} = 0
	& \textrm{in }\Omega_T
	\,,
\\\quad
{\bf u} = {\bf 0}
	& \textrm{on }\partial\Omega_T
	\,,
\\
\frac{\partial \phi}{\partial t}
	+{\rm Pe} ({\bf u}\cdot\nabla)\phi
    = \lambda{\rm div}\brkts{
    \nabla\brkts{
        f(\phi) 
        -\Delta\phi
        }
    }
    & \textrm{in }\Omega_T\,,
\\\quad
\nabla_n\phi:= {\bf n}\cdot\nabla\phi
    = g({\bf x})
    & \textrm{on }\partial\Omega_T
    \,,
\\\quad
\nabla_n\Delta\phi
    = 0
    & \textrm{on }\partial\Omega_T
    \,,
\\\quad
\phi({\bf x},0)
    = h({\bf x})
    & \textrm{on }\Omega\,,
\end{cases}
}{PhMo}
where $\Omega_T:=\Omega\times]0,T[$, $\partial\Omega_T:=\partial\Omega^1\times]0,T[$, $\lambda$ represents
the elastic relaxation time of the system, and the driving force $\pmb{\eta}$ accounts for the elastic energy \cite{Liu2003}
\bsplitl{
\pmb{\eta}
	= -\gamma{\rm div}\brkts{
		\nabla\phi\otimes\nabla\phi
	}\,,
}{ElEn}
where $\gamma$ corresponds to the surface tension
\cite{Liu2001}. As in \cite{Abels2011}, we will set $\gamma=\lambda$ for simplicity.
The dimensionless parameter ${\rm Pe}:=\frac{k\tau L{\rm U}}{D}$ is the P\'eclet
number for a reference fluid velocity ${\rm U}:=\av{{\bf u}}$, $L$ is the
characteristic length of the porous medium, and the diffusion constant
$D=k\tau M$ obtained from the mobility via Einstein's relation for the temperature $\tau$
and the Bolzmann constant $k$. We note that the
immiscible flow equations can immediately be written for the full
incompressible Navier-Stokes equations as in \cite{Liu2003}. Our restriction
to the Stokes equation is motivated here by the fact that such flows turn
into Darcy's law in porous media \cite{Carbonell1983,Hornung1997}.

\begin{figure}[htbp]
\begin{center}
\includegraphics[width=11.5cm]{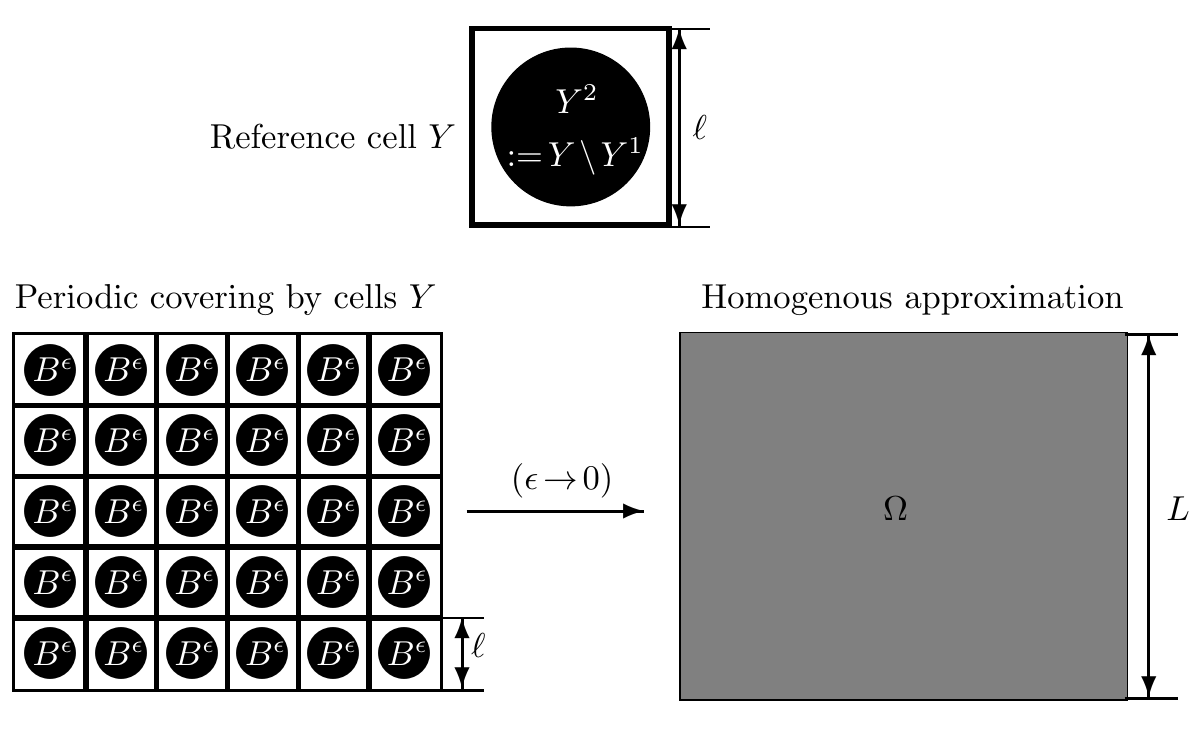}
\caption{{\bf Left:} Porous medium $\Omega^\epsilon:=\Omega\setminus B^\epsilon$ as a periodic covering of reference cells $Y:=[0,\ell]^d$.
{\bf Top:} Definition of the reference cell $Y=Y^1\cup Y^2$ with $\ell=1$.
{\bf Right:} The ``homogenization limit'' $\epsilon:=\frac{\ell}{L}\to 0$ scales the perforated domain such that perforations
become invisible in the macroscale.}
\label{fig:MicMac}
\end{center}
\end{figure}

\medskip

The main objective of our study is the derivation of effective macroscopic
equations describing \reff{PhMo} in the case of perforated domains
$\Omega^\epsilon\subset\mathbb{R}^d$ instead of a homogeneous
$\Omega\subset\mathbb{R}^d$. A useful and reasonable approach is to represent
a porous medium $\Omega=\Omega^\epsilon\cup B^\epsilon$ periodically with
pore space $\Omega^\epsilon$ and solid phase $B^\epsilon$. The arising
interface $\partial\Omega^\epsilon\cap\partial B^\epsilon$ is denoted by
$I^\epsilon$. As usual,  the dimensionless variable $\epsilon>0$ defines the
heterogeneity $\epsilon =\frac{\ell}{L}$ where $\ell$ represents the
characteristic pore size and $L$ is the characteristic length of the porous
medium, see Figure \ref{fig:MicMac}. The porous medium is defined by a
periodic coverage of a reference cell $Y:=
[0,\ell_1]\times[0,\ell_2]\times\dots\times[0,\ell_d]$, $\ell_i\in\mathbb{R}$, $i=1,\dots,d$, which represents a
single, characteristic pore. The periodicity assumption allows, by passing
to the limit $\epsilon\to 0$ (see Figure \ref{fig:MicMac}) for the
derivation effective macroscopic porous media equations.
The pore and the solid phase of the medium are defined as
usual by,
\bsplitl{ \Omega^\epsilon
    & := \bigcup_{{\bf z}\in\mathbb{Z}^d}\epsilon\brkts{Y^1+{\bf z}}\cap\Omega\,,
\qquad
B^\epsilon
    := \bigcup_{{\bf z}\in\mathbb{Z}^d}\epsilon\brkts{Y^2+{\bf z}}\cap\Omega
    =\Omega\setminus\Omega^\epsilon\,,
}{Oe2} where the subsets $Y^1,\,Y^2\subset Y$ are defined such that
$\Omega^\epsilon$ is a connected set. More precisely, $Y^1$ denotes the pore
phase (e.g. liquid or gas phase in wetting problems), see Figure
\ref{fig:MicMac}.

These definitions allow us to rewrite \reff{PhMo}
by the following microscopic formulation
\bsplitl{
\textrm{(Porous case)}\quad
\begin{cases}
-\epsilon^2\mu\Delta{\bf u}_\epsilon
	+\nabla p_\epsilon
	=  -\gamma{\rm div}\brkts{
		\nabla\phi_\epsilon\otimes\nabla\phi_\epsilon
	}
	& \textrm{in }\Omega_T^\epsilon
	\,,
\\\quad
{\rm div}\,{\bf u}_\epsilon = 0
	& \textrm{in }\Omega_T^\epsilon
	\,,
\\\quad
{\bf u}_\epsilon = {\bf 0}
	& \textrm{on }I_T^\epsilon
	\,,
\\
\pd{}{t}\phi_\epsilon
	+ {\rm Pe}({\bf u}_\epsilon\cdot\nabla)\phi_\epsilon
    = \lambda{\rm div}\brkts{
    \nabla\brkts{
        f(\phi_\epsilon) 
        -\Delta\phi_\epsilon
        }
    }
    & \textrm{in }\Omega_T^\epsilon\,,
\\\quad
\nabla_n\phi_\epsilon
	:= {\bf n}\cdot\nabla\phi_\epsilon
    	= g_\epsilon({\bf x})
	:= g({\bf x}/\epsilon)
    & \textrm{on }I_T^\epsilon
    \,,
\\\quad
\nabla_n\Delta\phi_\epsilon
    = 0
    & \textrm{on }I_T^\epsilon
    \,,
\\\quad
\phi_\epsilon(\cdot,0)
    = \psi(\cdot)
    & \textrm{on }\Omega^\epsilon\,,
\end{cases}
}{PeMoPr}
where $I_T^\epsilon:=I^\epsilon\times]0,T[$. $g_\epsilon({\bf x})=g({\bf x}/\epsilon)$ is now a periodic wetting
boundary condition accounting for the wetting properties of the pore walls. Even under the assumption
of periodicity, the microscopic system \reff{PeMoPr} leads to a high-dimensional
problem, since the space discretization parameter needs to be chosen to
be much smaller than the characteristic size $\epsilon$ of the
heterogeneities of the  porous structure, e.g. left-hand side of Figure
\ref{fig:MicMac}. The homogenization method
provides a systematic tool for reducing the intrinsic dimensional complexity by
reliably averaging over the microscale represented by a single periodic
reference pore $Y$. We note that the nonlinear nature of problem \reff{PeMoPr}
exploits a scale separation with respect to the upscaled chemical potential, see
Definition \ref{def:SSCP}, for the derivation of the effective macroscopic interfacial evolution
in strongly heterogeneous environments.  Such a scale separation
turns out to be the key for the upscaling/homogenization of nonlinear problems,
see \cite{Schmuck2011,Schmuck2012m,Schmuck2012b}.

Obviously, the systematic and reliable derivation of practical, convenient,
and low-dimensional approximations is the key to feasible numerics of
problems posed in porous media and provides a basis for computationally
efficient schemes. To this end, we relax the full microscopic formulation
\reff{PeMoPr} further by restricting \reff{PeMoPr} to periodic fluid flow. By
taking the stationary version of equation \reff{PeMoPr}$_4$ on to a single
periodic reference pore $Y$ and by denoting the according stationary solution
by $\Phi(\cdot)$, we can formulate the following periodic flow
problem
\bsplitl{
\textrm{(Periodic flow)}\quad
\begin{cases}
-\mu\Delta_{\bf y}{\bf u}
	+\nabla_{\bf y} p
	=  \pmb{\eta}
	& \textrm{in }Y^1
	\,,
\\\quad
{\rm div}_{\bf y}\,{\bf u} = 0
	& \textrm{in }Y^1
	\,,
\\\quad
{\bf u} = {\bf 0}
	& \textrm{on }\partial Y^2
	\,,
\\\quad
\textrm{${\bf u}$ is $Y^1$-periodic}\,,	
\\
{\rm Pe}({\bf u}\cdot\nabla_{\bf y})\Phi
	= \lambda{\rm div}_{\bf y}\brkts{
		\nabla_{\bf y}(f(\Phi)-\Delta_{\bf y}\Phi)
	}
	&\textrm{in }Y^1\,,
\\\quad
\nabla_n\Phi
	: = ({\bf n}\cdot\nabla_{\bf y})\Phi
	=g({\bf y})
	&\textrm{on }\partial Y^2\,,
\\\quad
\nabla_n\Delta_{\bf y}\Phi
	= 0
	&\textrm{on }\partial Y^2\,,
\\\quad
\textrm{$\psi$ is $Y^1$-periodic}\,.
\end{cases}
}{PeFl}
We remark that in certain occasions it might be suitable to further reduce problem
\reff{PeFl}. For instance, in general the reference cell is only filled
by one fluid phase, i.e., $\nabla_{\bf y}\Phi=0$ almost everywhere in $Y^1$, and hence one only needs to solve for
the periodic Stokes problem \reff{PeFl}$_1$--\reff{PeFl}$_4$
by replacing the self-induced driving force
\reff{ElEn}
with the constant driving force $\pmb{\eta}:={\bf e}_1$ where ${\bf e}_1$ denotes the canonical basis vector in the
${\bf x}_1$-direction of the Euclidean space. The periodic fluid velocity defined by \reff{PeFl} for
such an $\pmb{\eta}$ can be considered as the
spatially periodic velocity of a moving frame \cite{Allaire2010a}. Motivated by
\cite{Allaire2010a,Mei1992,Rubinstein1986}, we study the case of large P\'eclet number and
consider the following distinguished case:

\medskip

{\bf Assumption (LP):} \emph{The P\'eclet number scales with respect to the
characteristic pore size $\epsilon>0$ as follows: ${\rm
Pe}\sim\frac{1}{\epsilon}\,.$ }

\medskip

Let us first discuss Assumption {\rm (LP)}. If one introduces the microscopic
P\'eclet number ${\rm Pe}_{mic}:=\frac{k\tau\ell{\rm U}}{D}$, then it follows
immediately that ${\rm Pe}=\frac{{\rm Pe}_{mic}}{\epsilon}$. Since we
introduced a periodic flow problem on the characteristic length scale
$\ell>0$ of the pores by problem \reff{PeFl}, it is obvious that we have to
apply the microscopic P\'eclet number in a corresponding microscopic
formulation, see \reff{MiPr} below. Hence, the periodic fluid velocity ${\bf
u}({\bf x}/\epsilon):={\bf u}({\bf y})$ enters the microscopic phase field
problem as follows
\bsplitl{
\textrm{(Microscopic problem)}\quad
\begin{cases}
\pd{}{t}\phi_\epsilon
	+ \frac{{\rm Pe}_{mic}}{\epsilon}({\bf u}({\bf x}/\epsilon)\cdot\nabla)\phi_\epsilon
\\\qquad
     = \lambda{\rm div}\brkts{
    \nabla\brkts{
        f(\phi_\epsilon) 
        -\Delta\phi_\epsilon
        }
    }
    & \textrm{in }\Omega_T^\epsilon\,,
\\\quad
\nabla_n\phi_\epsilon
	:= {\bf n}\cdot\nabla\phi_\epsilon
    	= g({\bf x}/\epsilon)
    & \textrm{on }I_T^\epsilon
    \,,
\\\quad
\nabla_n\Delta\phi_\epsilon
    = 0
    & \textrm{on }I_T^\epsilon
    \,,
\\\quad
\phi_\epsilon(\cdot,0)
    = \psi(\cdot)
    & \textrm{on }\Omega^\epsilon\,.
\end{cases}
}{MiPr}
We note that with our subsequently applied upscaling strategy, we do not
account for boundary layers in the vicinity of rigid boundaries. Such
boundary layers become increasingly important in the case of large
P\'eclet numbers. Moreover, we make use of the splitting strategy
introduced in \cite{Schmuck2012a} and here extended to fluid flow, i.e.,
\bsplitl{
\textrm{(Splitting)}\quad
\begin{cases}
\quad
\frac{\partial}{\partial t}(-\Delta_\epsilon)^{-1}w_\epsilon
	+\frac{\rm Pe_{mic}}{\epsilon}\brkts{
		{\bf u}({\bf x}/\epsilon)\cdot\nabla
	}(-\Delta_\epsilon)^{-1}w_\epsilon
	&
\\\qquad\qquad
	= \lambda\brkts{
	{\rm div}\brkts{
		\hat{\rm M}\nabla w_\epsilon
	}
	+ {\rm div}\brkts{
		\hat{\rm M}\nabla f(\phi_\epsilon)
	}
	}
&
\textrm{in }\Omega_T^\epsilon\,,
\\\qquad
-\nabla_n\Delta\phi_\epsilon
    = \nabla_n w_\epsilon
    = 0
    &\textrm{on }I_T^\epsilon
    \,,
\\\quad
-\Delta_\epsilon\phi_\epsilon
	= w_\epsilon
\\\qquad
\nabla_n \phi_\epsilon
    	= g({\bf x}/\epsilon)
	=g_\epsilon({\bf x})
    &\textrm{on }I_T^\epsilon
    \,,
\\\quad
\phi_\epsilon(\cdot,0)
    = \psi(\cdot)
    &\textrm{on }\Omega^\epsilon\,,
\end{cases}
}{SpSt}
where we will properly define $\Delta_\epsilon={\cg A}_\epsilon$ in Section \ref{sec:Proof}.

The main result of our study is the systematic derivation of upscaled
immiscible flow equations which effectively account for the pore geometry
starting from the microscopic system \reff{PeFl}--\reff{MiPr}, i.e.,
\bsplitl{ \textrm{(Upscaled equation)}\quad
\begin{cases}
\quad
p\pd{\phi_0}{t}
	-{\rm div}\Bigl(
		\hat{\rm C}\nabla \phi_0
	\Bigr)
    = \lambda{\rm div}\Bigl(
		\hat{\rm M}_\phi\nabla f(\phi_0)
	\Bigr)
\\\qquad\quad
    -\frac{\lambda}{p}{\rm div}\brkts{
        \hat{\rm M}_w\nabla \brkts{
            {\rm div}\brkts{
                \hat{\rm D}\nabla \phi_0
            }
            -\tilde{g}_0
        }
    }
\,,
\end{cases}
}{EffImFl}
where $\hat{\rm C}$ takes the fluid convection
into account. These two tensors account for the so-called
diffusion-dispersion relations (e.g. Taylor-Aris-dispersion
\cite{Aris1956,Brenner1980,Taylor1953}). The result \reff{EffImFl} makes use
of the recently proposed splitting strategy for the homogenization of fourth
order problems in \cite{Schmuck2012a} and an asymptotic multiscale expansion
with drift (i.e., moving frame) introduced in \cite{Allaire2010a,Marusic-Paloka2005}.

The manuscript is organized as follows. We present our main
results in Section \ref{sec:MaRe} and the corresponding proofs follow in the subsequent Section
\ref{sec:Proof}. Concluding remarks and open questions are offered in Section
\ref{sec:Cncls}.

%

\csection{Preliminaries and notation}\label{sec:Prelim}
We recall basic results required for our subsequent analysis which depends also on 
certain notational conventions. We consider connected macroscopic domains $\Omega$ with 
Lipschitz continuous boundaries $\partial\Omega$. Under the usual conventions for Sobolev spaces, 
we say that $u\in W^{k,p}(\Omega)$ if and only if
\bsplitl{
\N{u}{k,p}^p
	:= \sum_{\av{\alpha}\leq k} \N{D^\alpha u}{L^p}^p
	<\infty
}{Wkp}
for a multi-index $\alpha$ such that $D^\alpha:=\frac{\partial^{\av{\alpha}}}{\partial^{\alpha_1}x_1\dots\partial^{\alpha_d}x_d}$ and $p<\infty$. Herewith, we can identify corresponding Hilbert spaces ($p=2$) by 
$H^k(\Omega):=W^{k,2}(\Omega)$.
We introduce the following (energy) space of functions
\bsplitl{
H^2_E(\Omega)
	:=\brcs{
		v\in H^2(\Omega)\,\bigl|\,\nabla_nv=0\quad\textrm{on}\quad\partial\Omega
	}\,,
}{H2E}
which naturally appears in the context of weak solutions for the phase field equations 
\reff{PhMo}$_4$. In order to account for the periodic reference cells appearing 
due to asymptotic multiscale-expansions/homogenization, we define $\overline{H}^1_{per}(Y)$ 
as the closure of $C^\infty_{per}(Y)$ in the $H^1$-norm where $C^\infty_{per}(Y)$ is 
the subset of $Y$-periodic functions of $C^\infty(\mathbb{R}^d)$. As we need uniqueness 
of solutions, we will work with the following space of functions
\bsplitl{
H^1_{per}(Y)
	:= \brcs{
		u\in\overline{H}^1_{per}(Y)\, \bigl|\, {\cg M}_Y(u)=0
	}
	\,,
}{H1per}
where ${\cg M}_Y(u):=\frac{1}{\av{Y}}\int_Yu\,d{\bf y}$. 
%
%
%
%
%
%
In order to establish the existence and uniqueness of weak solutions of the upscaled convective phase field 
equations, we need the following Aubin-Lions compactness result (e.g. \cite{Showalter1997}), i.e.,

\begin{thm}\label{thm:AL}\emph{(Aubin-Lions)} Let $X_0,\,X,\,X_1$ be Banach spaces with 
$X_0\subset X\subset X_1$ and assume that $X_0\hookrightarrow X$ is compact and 
$X\hookrightarrow X_1$ is continuous. Let $1<p<\infty$, $1<q<\infty$ and let $X_0$ and 
$X_1$ be reflexive. Then, for $W:=\brcs{u\in L^p(0,T;X_0)\,\bigr|\,\partial_tu\in L^q(0,T;X_1)}$ the 
inclusion $W\hookrightarrow L^p(0,T;X)$ is compact.
\end{thm}

\csection{Results: Effective immiscible flow equations in porous media}\label{sec:MaRe}
The presentation of our main result depends on the following:

\medskip

\begin{defn}\label{def:SSCP} \emph{(Scale separation)} We say that the the macroscopic chemical potential is scale
separated if and only if the upscaled chemical potential
\bsplitl{
\mu_0
	:=
	f(\phi)-\Delta\phi
	\,,
}{ChPo}
satisfies $\frac{\partial \mu_0}{\partial x_l}=0$ for each $1\leq l\leq d$ on the level of the reference cell $Y$ but not 
in the macroscopic domain $\Omega$.
\end{defn}

\medskip

\begin{rem}\label{rem:ScSe} We note that the scale separation in Definition \ref{def:SSCP} follows intuitively 
from the key requirement in homogenization theory that one can identify a slow (macroscopic) variable and 
at least one fast (microscopic) variable. Hence, the above scale separation means that the macroscopic variable does 
not vary over the dimension of the microscale defined by a characteristic reference cell.
\end{rem}

The scale separation \reff{def:SSCP} emerges as a key requirement for the homogenization of
nonlinearly coupled partial differential equations 
in order to guarantee the mathematical well-posedness of
the corresponding cell problems which define effective transport coefficients
in homogenized, nonlinear (and coupled) problems \cite{Schmuck2011,Schmuck2012m,Schmuck2012b}.

We note that the upscaling requires to identify effective macroscopic boundary conditions on the 
macroscopic domain $\Omega$. Such a condition will be denoted by $\tilde{h}_0$ below. In fact, $\tilde{h}_0$ 
can be computed as $\tilde{g}_0$. We summarize our main result in the following

\begin{thm}\label{thm:Main} \emph{(Effective convective Cahn-Hilliard equation)}
We assume that the \emph{Assumption (LP)} holds and that the macroscopic chemical potential $\mu_0$ 
satisfies the scale separation property in the sense of \emph{Definition \ref{def:SSCP}} and let $\psi({\bf x})\in H^2(\Omega)$.

Then, the microscopic equations \reff{PeFl}--\reff{MiPr} for immiscible flow in porous media
admit the following effective macroscopic form after averaging over the microscale, i.e.,
\bsplitl{
\begin{cases}
\quad
p\pd{\phi_0}{t}
	-{\rm div}\Bigl(
		\hat{\rm C}\nabla \phi_0
	\Bigr)
    = \lambda{\rm div}\Bigl(
 		\hat{\rm M}_\phi\nabla f(\phi_0)
	\Bigr)
\\\qquad\quad
    -\frac{\lambda}{p}{\rm div}\brkts{
        \hat{\rm M}_w\nabla \brkts{
            {\rm div}\brkts{
                \hat{\rm D}\nabla \phi_0
            }
            -\tilde{g}_0
        }
    }
	& \textrm{in }\Omega_T
\,,
\\\quad
\nabla_n\phi_0
	:= {\bf n}\cdot\nabla\phi_0
    	= \tilde{h}_0({\bf x})
    & \textrm{on }\partial\Omega\times]0,T[
    \,,
\\\quad
\nabla_n\Delta\phi_0
    = 0
    & \textrm{on }\partial\Omega\times]0,T[
    \,,
\\\quad
\phi_0(\cdot,0)
    = \psi(\cdot)
    & \textrm{in }\Omega
	\,,
\end{cases}
}{EffImFlThm}
where the tensors $\hat{\rm C}:=\brcs{{\rm c}_{ik}}_{1\leq i,k\leq d}$,
$\hat{\rm D}:=\brcs{{\rm d}_{ik}}_{1\leq i,k\leq d}$,
$\hat{\rm M}_\phi = \brcs{{\rm m}^\phi_{ik}}_{1\leq i,k\leq d}$, and
$\hat{\rm M}_w=\brcs{{\rm m}^w_{ik}}_{1\leq i,k\leq d}$
are defined by
\bsplitl{
\begin{cases}
\quad
{\rm c}_{ik}
	:= \frac{{\rm Pe}_{mic}}{\av{Y}}\int_{Y^1}({\rm u}^i-{\rm v}^i)
			\delta_{ik}\xi^k_\phi
	\,d{\bf y}\,,
	&
\\\quad
{\rm d}_{ik}
    := \frac{1}{\av{Y}}\sum^d_{j=1}\int_{Y^1}\brkts{
        \delta_{ik} - \delta_{ij}\pd{\xi^k_\phi}{y_j}
        }
    \,d{\bf y}\,,
	&
\\\quad
{\rm m}^\phi_{ik}
    :=
    \frac{1}{\av{Y}}\sum_{j=1}^d\int_{Y^1}\brkts{
        {\rm m}_{ik}
        -{\rm m}_{ij}\pd{\xi^k_\phi}{y_j}
    }\,d{\bf y}\,,
	&
\\\quad
{\rm m}^w_{ik}
    :=
    \frac{1}{\av{Y}}\sum_{j=1}^d\int_{Y^1}\brkts{
        {\rm m}_{ik}
        -{\rm m}_{ij}\pd{\xi^k_w}{y_j}
    }\,d{\bf y}\,.
	&
\end{cases}
}{kappaThm} 
The effective fluid velocity ${\bf v}$ is defined by ${\rm
v}^j:=\frac{\rm Pe_{mic}}{\av{Y^1}}\int_{Y^1}{\rm u}^j({\bf y})\,d{\bf y}$ where {\bf u}
solves the periodic reference cell problem \reff{PeFl}. The effective 
wetting boundary condition on the pore walls becomes 
$\tilde{g}_0:=-\frac{\gamma}{C_h}\frac{1}{\av{Y}}\int_{\partial Y^1}
        \brkts{
            a_1\chi_{\partial Y^{1}_{w}}({\bf y})
            +a_2\chi_{\partial Y^{2}_{w}}({\bf y})
        }
    \,d{\bf y}$ 
and on the boundary $\partial\Omega$ of the macroscopic domain 
$\Omega$ we impose 
$\tilde{h}_0:=-\frac{\gamma}{C_h}\frac{1}{\av{Y}}\int_\Gamma
        \brkts{
            a_\Gamma({\bf y})
        }
    \,d{\bf y}$. 
The corrector
functions $\xi^k_\phi\in H^1_{per}(Y^1)$ and $\xi^k_w\in
L^2(\Omega;H^1_{per}(Y^1))$ for $1\leq k,l\leq d$ solve in the distributional
sense the
following reference cell problems
\bsplitl{
\xi_\phi^k:\quad
\begin{cases}
-\sum_{i,j=1}^d
    \pd{}{y_i}\brkts{
        \delta_{ik}-\delta_{ij}\pd{\xi^k_\phi}{y_j}
    }
    = 0
    &\textrm{ in }Y^1\,,
\\
\sum_{i,j=1}^d{\rm n}_i
	\brkts{
        \delta_{ij}\pd{\xi^k_\phi}{y_j}
        -\delta_{ik}
        }
    =
    0
    &\textrm{ on }\partial Y^1\,,
\\
\xi^k_\phi({\bf y})\textrm{ is $Y$-periodic and ${\cg M}_{Y^1}(\xi^k_\phi)=0$,}
\end{cases}
}{xivTh}
\bsplitl{
\xi_w^k:\quad
\begin{cases}
-\sum_{i,j=1}^d
    \pd{}{y_i}\brkts{
        \delta_{ik}-\delta_{ij}\pd{\xi^k_w}{y_j}
    }
\\\qquad\qquad
    =
    \lambda\biggl( \sum_{i,j=1}^d\pd{}{y_i}\brkts{
        {\rm m}_{ik}
        -{\rm m}_{ij}\pd{\xi^k_\phi}{y_j}
    }
    &\textrm{ in }Y^1\,,
\\
\sum_{i,j=1}^d{\rm n}_i\biggl( 
        \brkts{
        \delta_{ij}\pd{\xi^k_w}{y_j}
        -\delta_{ik}
        }
\\\qquad\qquad
        -\lambda\sum_{i,j=1}^d\pd{}{y_i}\brkts{
            {\rm m}_{ik}
            -{\rm m}_{ij}\pd{\xi^k_\phi}{y_j}
        }
    = 0
    &\textrm{ on }\partial Y^1\,,
\\
\xi^k_w({\bf y})\textrm{ is $Y$-periodic and ${\cg M}_{Y^1}(\xi^k_w)=0$.}
\end{cases}
}{xiwThm}
\end{thm}

\medskip

\begin{rem}\label{rem:xiphixiw} \emph{(Isotropic mobility)} The cell problem \reff{xiwThm} is equal to problem 
\reff{xivTh} if we consider the case of isotropic mobility tensors, i.e., $\hat{\rm M}:=m\hat{\rm I}$. In this special 
case, we immediately have $\xi_\phi^k=\xi_w^k$ and hence the porous media correction tensors satisfy 
$m\hat{\rm D}=\hat{\rm M}_\phi=\hat{\rm M}_w$.
\end{rem}

\medskip

The next theorem guarantees the well-posedness of \reff{UpScPrTh} in the sense 
of week solutions. For convenience, we achieve existence of weak solutions for polynomial 
free energies in the sense of Assumption (PF) \cite{Temam1997}.

\medskip

\begin{thm}\label{thm:2} \emph{(Existence \& Uniqueness)} Let $\psi\in L^2(\Omega)$, $T^*>0$, and 
assume that the admissible free energy densities $F$ in \reff{FrEn} satisfy \emph{Assumption (PF)}. Then, there 
exists a unique solution 
$\phi_0\in L^\infty(]0,T^*[;L^2(\Omega))\cap L^2([0,T^*[;H^2_E(\Omega))$ 
to the following upscaled/homogenized problem 
\bsplitl{
\begin{cases}
\quad
p\frac{\partial \phi_0}{\partial t}
	+\frac{\lambda}{p}{\rm div}\brkts{
		\hat{\rm M}_w\nabla\brcs{
			{\rm div}\brkts{
				\hat{\rm D}\nabla\phi_0
			}
			-\tilde{g}_0
		}
	}
	&
\\\qquad\quad
	= {\rm div}\brkts{
		\ebrkts{
			\lambda f'(\phi_0)\hat{\rm M}_\phi
			+\hat{\rm C}
		}\nabla\phi_0
	}
	&\textrm{in }\Omega_T\,,
\\\quad
\nabla_n\phi_0
	=0
	&\textrm{on }\partial\Omega\times]0,T^*[\,,
\\\quad
\nabla_n\Delta\phi_0
	=0
	&\textrm{on }\partial\Omega\times]0,T^*[\,,
\\\quad
\phi_0(\cdot,0)
	= \psi(\cdot)
	&\textrm{in }\Omega\,.
\end{cases}
}{UpScPrTh}
\end{thm}

\medskip

We prove Theorem \ref{thm:2} by adapting arguments from \cite[Section 2, p.151]{Temam1997} to 
the homogenized setting including the diffusion-dispersion tensor which 
accounts for periodic flow.
\subsection{Numerical computations}

\begin{figure}[htbp]
\begin{center}
\includegraphics[width=11.5cm]{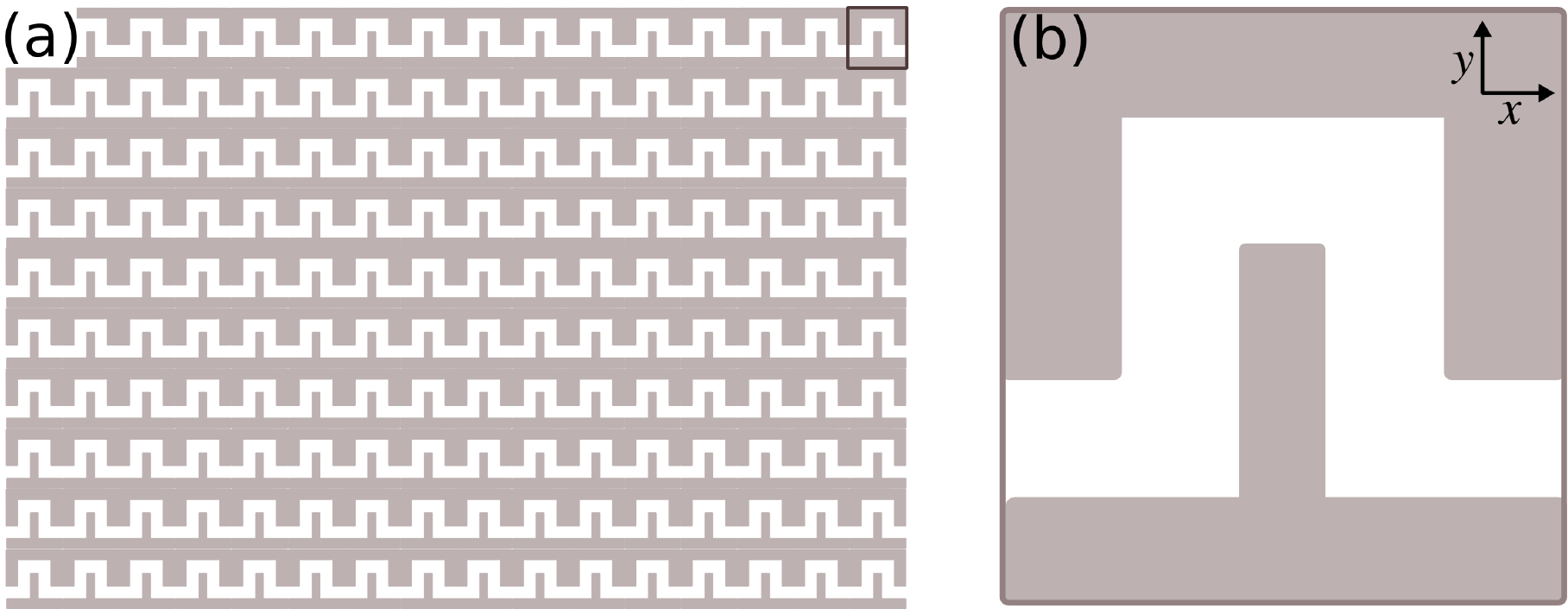}
\caption{(a): Example of the 2D porous medium considered for the numerical computations. The reference
cell consists of a non-straight channel of constant cross-section which is periodic in the
$x$ direction as depicted in (b). Gray area corresponds to the solid phase of the medium.}
\label{fig:NumPorous}
\end{center}
\end{figure}

To exemplify the results presented above we perform a numerical study of the effective
macroscopic Cahn-Hilliard equation \reff{EffImFlThm}. We consider a two-dimensional
(2D) porous medium consisting of a series of periodic reference cells the geometrical shape
of which is a non-straight channel of constant cross-section (see Figure \ref{fig:NumPorous})
with periodic boundary conditions at the inlet and outlet. The Cartesian coordinates in
the microscopic problem are named as $x$ and $y$ which correspond to the $y_1$ and
$y_2$ variables used in the definition of the domain $Y^1$, respectively. We define
the geometry in such a way that the porosity of the medium is $p=0.46$.
The macroscopic domain $\Omega$ is compound of $35$ reference cells in the perpendicular
direction $y$ of the flow and $50$ in the $x$ direction.  We fix a constant driving
force $U=1$ at the inlet of the system by fixing the gradient of the chemical
potential\cite{Pradas2012}. For simplicity we take the macroscopic mobility $M=1$
which gives rise to a microscopic P{\'e}clet number ${\rm Pe}_{mic}= 0.04$.

\begin{figure}[htbp]
\begin{center}
\includegraphics[width=11.5cm]{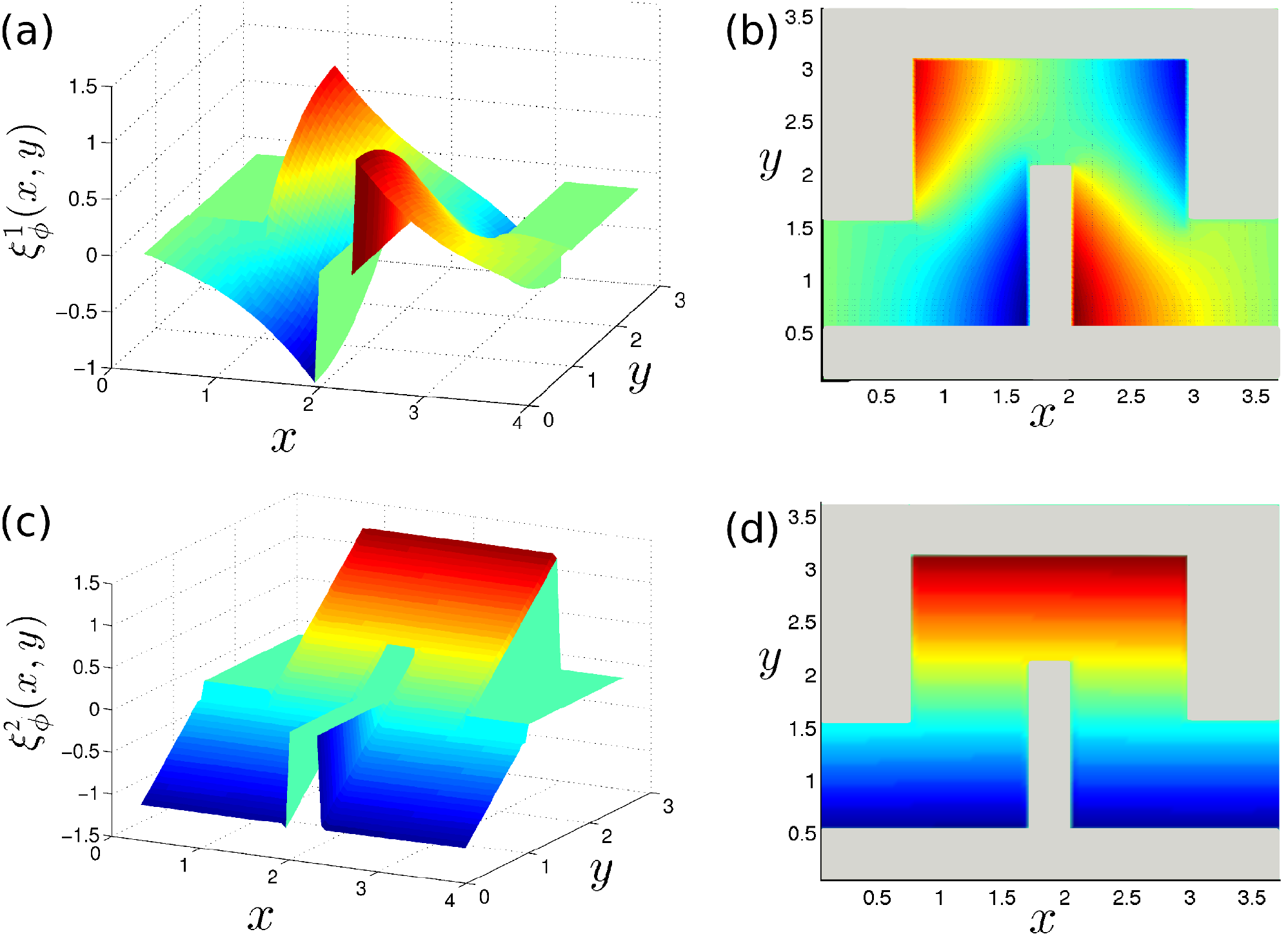}
\caption{Corrector functions  $\xi_\phi^1(x,y)$ (a,b) and  $\xi_\phi^2(x,y)$ (c,d) for
the particular reference cell defined in Figure \ref{fig:NumPorous}b. Panels (a) and
(c) show the three-dimensional plots, and panels (b) and (d) show the corresponding 2D
projection onto the plane $(x,y)$.}
\label{fig:NumXi}
\end{center}
\end{figure}

We first compute the components of the different tensors $\hat{\rm C}$, $\hat{\rm D}$,
$\hat{\rm M}_\phi$, and $\hat{\rm M}_w$ for which we need to solve the reference cell
problems \reff{xivTh} and \reff{xiwThm}. We consider the case of isotropic
mobility with $m=1$ and hence we have
$\xi_\phi^k=\xi_w^k$ and $\hat{\rm D}=\hat{\rm M}_\phi=\hat{\rm M}_w$.
In this case, the reference cell problem is reduced to:
\bsplitl{
\xi_\phi^k:\quad
\begin{cases}
\big(\frac{\partial^2}{\partial x^2}+
\frac{\partial^2}{\partial y^2}\big)\xi^k_\phi
    = 0
    &\textrm{ in }Y^1\,,
\\
\big({\rm n}_1\frac{\partial}{\partial x}+
{\rm n}_2\frac{\partial}{\partial y}\big)\xi^k_\phi
    =
    {\rm n}_k
    &\textrm{ on }\partial Y^1\,,
\end{cases}
}{NumxivTh}
which corresponds to the Laplace equation with special boundary conditions. The above
equation is solved by using a finite differences numerical scheme and the resulting
corrector functions $\xi_\phi^1(x,y)$ and $\xi_\phi^2(x,y)$ are plotted in Figure \ref{fig:NumXi}.

\begin{figure}[htbp]
\begin{center}
\includegraphics[width=11.5cm]{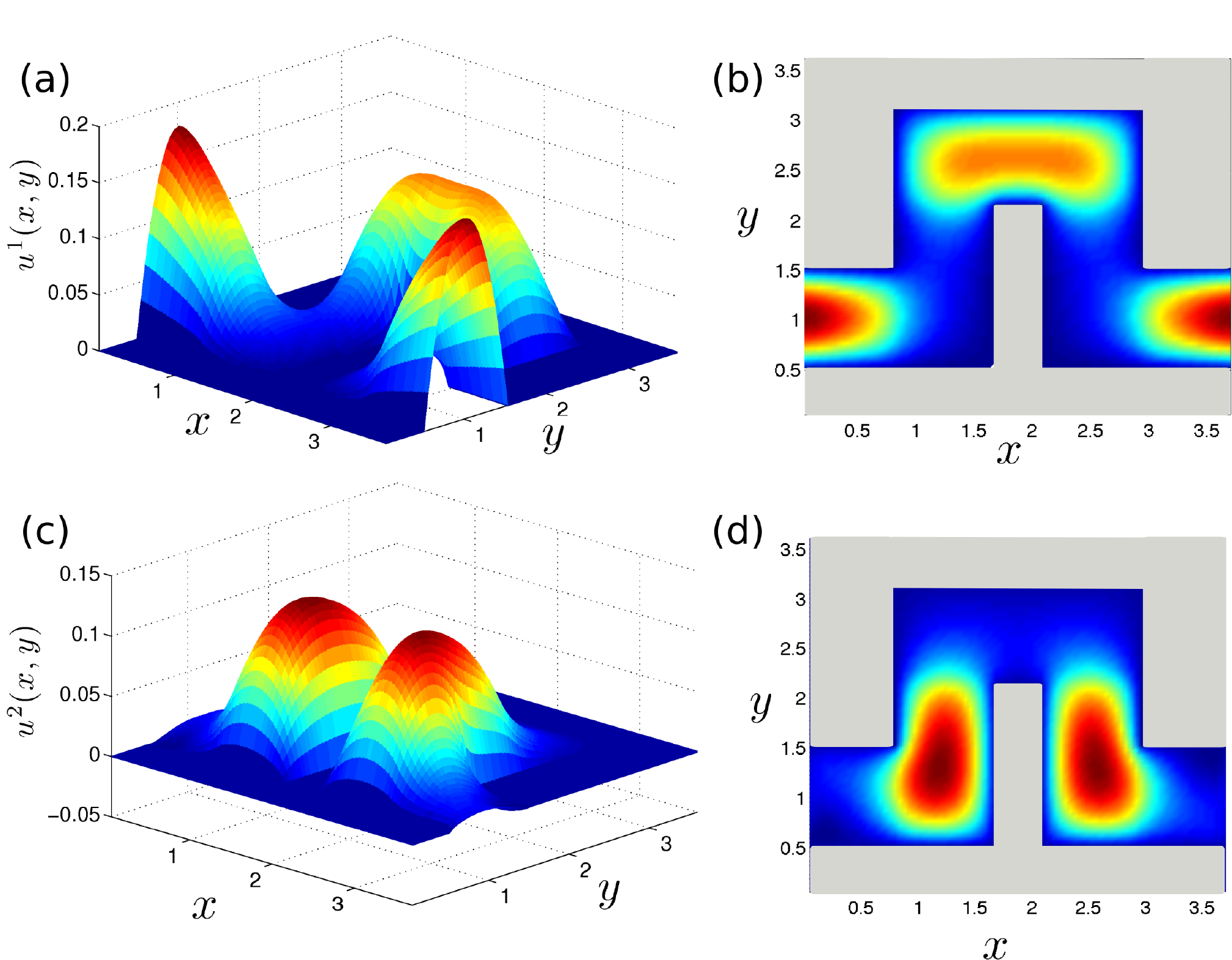}
\caption{Two components of the velocity field ${\rm u}^1(x,y)$ (a,b) and ${\rm u}^2(x,y)$ (c,d).
Panels (a) and (c) show the three-dimensional plots, and panels (b) and (d) show the
corresponding 2D projection onto the plane $(x,y)$.}
\label{fig:velflow}
\end{center}
\end{figure}

Once we know the corrector functions of the reference cell problem, we can compute
the different elements of the tensor $\hat{\rm D}$ as defined in \reff{kappaThm}
obtaining the values ${\rm d}_{11}=0.4$, and ${\rm d}_{12}={\rm d}_{21}={\rm d}_{22}=0$. Note that these
are similar values to those reported in \cite{Auriault1997} for a similar porous
geometry. Next we solve the Stokes flow for this particular microscopic
geometry by numerically integrating the periodic reference cell problem \reff{PeFl}
to obtain the velocity ${\bf u}$ and hence the coefficients for the tensor
$\hat{\rm C}$. The results for the two velocity components ${\rm u}^1(x,y)$ and ${\rm u}^2(x,y)$
are presented in Figure \ref{fig:velflow}. By applying the formula given in \reff{kappaThm}
we obtain the coefficients ${\rm c}_{11}=0.015$ and ${\rm c}_{22}=0.023$. Note that by definition
${\rm c}_{12}={\rm c}_{21}=0$.

\begin{figure}[htbp]
\begin{center}
\includegraphics[width=11.5cm]{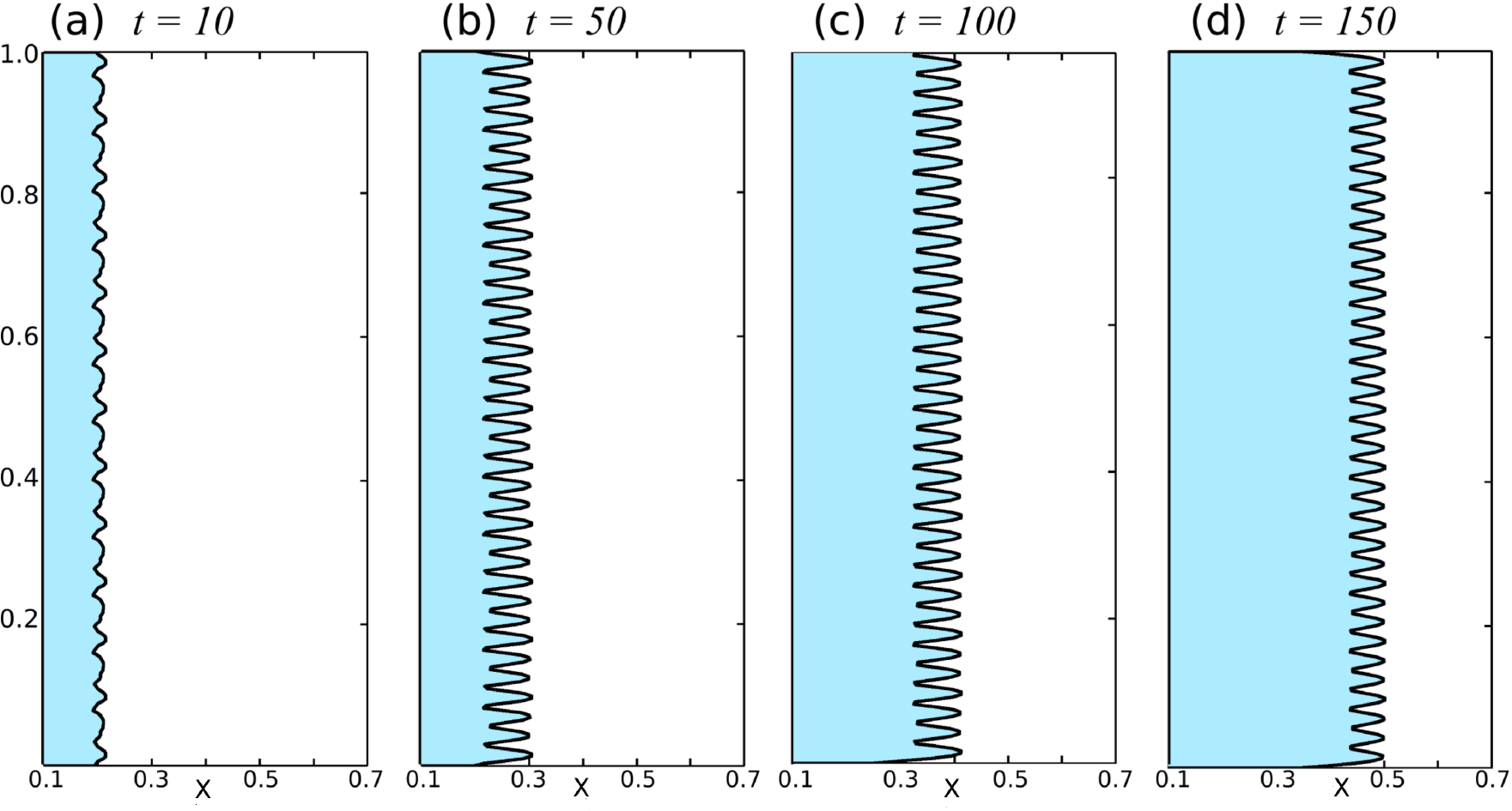}
\caption{Numerical integrations of the effective macroscopic Cahn-Hilliard equation at
different times. Blue color represents the liquid phase and the interface position is depicted as
a solid black line.}
\label{fig:pfmSim}
\end{center}
\end{figure}

Finally, with all the different tensor coefficients we can numerically
integrate the problem \reff{EffImFlThm} in the macroscopic domain 
$\Omega$, the Cartesian coordinates of which are denoted as $(X,Y)$.
We use a finite difference scheme for the spatial discretization and a
fourth-order Runge-Kutta algorithm with adaptive time stepsize to march
froward in time. The domain is discretized with a grid spacing $\Delta
X=0.01$ and we impose periodic boundary conditions along the transversal
direction of the flow. As an initial condition, we consider a small
sinusoidal shape for the interface separating the liquid from the gas phase.
Also, to simulate the same condition as in the porous medium, we impose the
driving force $U$ to be fixed alternately at the inlet of the system in such
a way it follows the periodicity of microscopic structure. The evolution of
the interface position is then found by setting $\phi_0(X,Y,t)=0$. The
results are presented in Figure \ref{fig:pfmSim} where we observe that the
profile of the interface evolves into a well defined spatial periodic shape
which corresponds to the periodic porous medium that is defined at the
microscopic level (had the macroscopic model ignored the microscopic details,
by e.g. taking the tensors to be identity matrices, the interface would be
flat at all times). For large times and after the influence of the initial
disturbances dies out, the interface approaches a steady travelling front
with a microstructure that reflects the porous medium structure (as
expected). Our results hence show that the effective macroscopic equation is
able to retain the microscopic details even though we do not resolve them
numerically.

\csection{Formal derivation of Theorem \ref{thm:Main}}\label{sec:Proof}
As in \cite{Schmuck2012a}, we introduce the differential operators
\bsplitl{
\begin{array}{ll}
{\cg A}_0
	=
	-\sum_{i,j=1}^d\pd{}{y_i}\brkts{\delta_{ij} \pd{}{y_j}}
	\,,
&\quad
{\cg B}_0
	=
	- \sum_{i,j=1}^d\pd{}{y_i}\brkts{{\rm m}_{ij}\pd{}{y_j}}
	\,,
\\
{\cg A}_1
	=
	-\sum_{i,j=1}^d\biggl[\pd{}{x_i}\brkts{\delta_{ij}\pd{}{y_j}}
&\quad
{\cg B}_1
	=
	-\sum_{i,j=1}^d\biggl[\pd{}{x_i}\brkts{{\rm m}_{ij}\pd{}{y_j}}
\\\quad\quad
		+\pd{}{y_i}\brkts{\delta_{ij}\pd{}{x_j}}
		\biggr]\,,
&\qquad\quad
		+\pd{}{y_i}\brkts{{\rm m}_{ij}\pd{}{x_j}}
		\biggr]\,,
\\
{\cg A}_2
	= - \sum_{i,j=1}^d\pd{}{x_j}\brkts{\delta_{ij}\pd{}{x_j}}\,,
&\quad
{\cg B}_2
	= - \sum_{i,j=1}^d\pd{}{x_j}\brkts{{\rm m}_{ij}\pd{}{x_j}}\,,
\end{array}
}{B0B1B2} which make use of the micro-scale $\frac{{\bf x}}{\epsilon}=:{\bf
y}\in Y$ such that ${\cg A}_\epsilon := \epsilon^{-2}{\cg A}_0 +
\epsilon^{-1}{\cg A}_1 +{\cg A}_2$, and ${\cg B}_\epsilon:=\epsilon^{-2}{\cg
B}_0 + \epsilon^{-1}{\cg B}_1 +{\cg B}_2$. Herewith, the Laplace operators
$\Delta$ and ${\rm div}\brkts{\hat{\rm M}\nabla}$ become $\Delta
u^\epsilon({\bf x}) = {\cg A}_\epsilon u({\bf x}, {\bf y})$ and ${\rm
div}\brkts{\hat{\rm M}\nabla}u^\epsilon({\bf x}) = {\cg B}_\epsilon u({\bf
x}, {\bf y})$, respectively, where $u^\epsilon({\bf x},t):=u({\bf
x}-\frac{{\bf v}}{\epsilon}t, {\bf y},t)$. Due to the drift \cite{Allaire2010a,Marusic-Paloka2005}, we additionally
have 
\bsplitl{
\frac{\partial}{\partial t} u^\epsilon
	= \brkts{\frac{\partial}{\partial t}-\frac{{\bf v}\cdot\nabla_{\bf x}}{\epsilon}}u^\epsilon
	\,, 
}{TiDe} 
where we find below by the Fredholm
alternative (or a solvability constraint) that ${\bf v}:=\frac{\rm
Pe_{loc}}{\av{Y^1}}\int_{Y^1}{\rm u}^j(y)\,dy\,.$ As in \cite{Schmuck2012a}
we apply the method of formal asymptotic multiscale expansions, that is,
\bsplitl{
w^\epsilon
	& := w_0({\bf x},{\bf y},t)
	+\epsilon w_1({\bf x},{\bf y},t)
	+\epsilon^2 w_2({\bf x},{\bf y},t)
	+\dots
	\,,
\\
\phi^\epsilon
	& := \phi_0({\bf x},{\bf y},t)
	+\epsilon \phi_1({\bf x},{\bf y},t)
	+\epsilon^2 \phi_2({\bf x},{\bf y},t)
	+\dots
	\,,
}{AsMuEx}
together with the splitting strategy introduced therein. In order to cope with the nonlinear 
form of the homogeneous free energy $f=F'$, see \reff{PNfrEn} and \reff{f}, we make use of a Taylor expansion 
which naturally leads to an expansion in $\epsilon$, i.e., 
\bsplitl{
f(\phi^\epsilon)
	= f(\phi_0)
	+f'(\phi_0)(\phi^\epsilon-\phi_0)
	+\frac{1}{2}f''(\phi_0)(\phi^\epsilon-\phi_0)^2
	+{\cal O}((\phi^\epsilon-\phi_0)^3)
	\,.
}{NlTlEp}
As a consequence, we obtain the
following sequence of problems by comparing terms of the same order in
$\epsilon$, with the first three problems being, \bsplitl{
\mathcal{O}(\epsilon^{-2}):\quad
\begin{cases}
\lambda\mathcal{B}_0\ebrkts{
		w_0
		+f(\phi_0)
	}
	+{\rm Pe}_{mic}\brkts{{\bf u}\cdot\nabla)_{\bf y}}{\cg A}_2^{-1}w_0
    = 0 
    &\textrm{in }Y^1\,,
\\\quad
\textrm{no flux b.c.}\,,
\\\quad
\textrm{$w_0$ is $Y^1$-periodic}\,,
\\
\mathcal{A}_0\phi_0=0
    &\textrm{in }Y^1\,,
\\\quad
\nabla_n \phi_0
    = 0
    &\textrm{on }\partial Y^1_w\cap\partial Y^2_w\,,
\\\quad
\textrm{$\phi_0$ is $Y^1$-periodic}\,,
\end{cases}
}{O-2n}
\bsplitl{
\mathcal{O}(\epsilon^{-1}):\quad
\begin{cases}
\lambda{\cg B}_0 \ebrkts{
		w_1
		+f'(\phi_0)\phi_1
	}
	+{\rm Pe}_{mic}\brkts{{\bf u}\cdot\nabla_{\bf y}}{\cg A}_2^{-1}w_1
\\\qquad
    = -\lambda{\cg B}_1 \ebrkts{
		w_0
		+f(\phi_0)
	}
	&
\\\qquad\quad
    -{\rm Pe}_{mic}\brkts{({\bf u}-{\bf v})\cdot\nabla}{\cg A}_2^{-1}w_0
    &\textrm{in }Y^1\,,
\\\quad
\textrm{no flux b.c.}\,,
\\\quad
\textrm{$w_1$ is $Y^1$-periodic}\,,
\\
{\cg A}_0 \phi_1
    = -{\cg A}_1 \phi_0
    &\textrm{in }Y^1\,,
\\\quad
\nabla_n \phi_1
    = 0
    &\textrm{on }\partial Y^1_w\cap\partial Y^2_w\,,
\\\quad
\textrm{$\phi_1$ is $Y^1$-periodic}\,,
\end{cases}
}{O-1n}
\bsplitl{
\mathcal{O}(\epsilon^{0}):\quad
\begin{cases}
\lambda{\cg B}_0\ebrkts{
		w_2
		+\frac{1}{2}f''(\phi_0)\phi_1^2
		+f'(\phi_0)\phi_2
	}
	+{\rm Pe}_{mic}\brkts{{\bf u}\cdot\nabla_{\bf y}}{\cg A}_2^{-1}w_2
\\\qquad
    = \lambda\brkts{
        {\cg B}_2\ebrkts{
		w_0
		+f(\phi_0)
	}
        +{\cg B}_1\ebrkts{
		w_1
		+f(\phi_0)\phi_1
	}
    }
&
\\
\qquad
	-{\rm Pe}_{mic}\brkts{({\bf u}-{\bf v})\cdot\nabla}{\cg A}_2^{-1}w_1
\\
\qquad
    -\partial_t{\cg A}_2^{-1}w_0
    &\textrm{in }Y^1\,,
\\\quad
\textrm{no flux b.c.}\,,
\\\quad
\textrm{$w_2$ is $Y^1$-periodic}\,,
\\
{\cg A}_0 \phi_2
    = -{\cg A}_2 \phi_0
    -{\cg A}_1 \phi_1
    +w_0
    &\textrm{in }Y^1\,,
\\\quad
\nabla_n \phi_2
    = g({\bf y})
    &\textrm{on }\partial Y^1_w\cap\partial Y^2_w\,,
\\\quad
\textrm{$\phi_2$ is $Y^1$-periodic}\,,
\end{cases}
}{O-0n}
As usual, the first problem \reff{O-2n} induces that the leading order terms $\phi_0$ and $w_0$
are independent of the micro-scale ${\bf y}$. The second problem \reff{O-1n} reads in explicit form for $\phi_1$ as
follows,
\bsplitl{
\xi_\phi:\quad
\begin{cases}
-\sum_{i,j=1}^d
    \pd{}{y_i}\brkts{
        \delta_{ik}-\delta_{ij}\pd{\xi^k_\phi}{y_j}
    }
    =
    &
\\\qquad\quad\,\,
    = -{\rm div}\brkts{
        {\bf e}_k-\nabla_y\xi^k_\phi
    }=0
    &\textrm{ in }Y^1\,,
\\
        {\bf n}\cdot\brkts{
            \nabla\xi^k_\phi
            +{\bf e}_k
        }
    =
    0
    &\textrm{ on }\partial Y^1_w\cap\partial Y^2_w\,,
\\
\xi^k_\phi({\bf y})\textrm{ is $Y$-periodic and ${\cg M}_{Y^1}(\xi^k_\phi)=0$,}
\end{cases}
}{Xiphi}
which represents the reference cell problem for $\phi_0$ after identifying
$\phi_1=-\sum_{k=1}^{d}\xi_\phi^k({\bf y})\frac{\partial \phi_0}{\partial x_k}$.

\medskip

The cell problem for $w_1$ is substantially more involved since it depends on
the fluid velocity ${\bf u}$ and the the corrector $\xi_\phi^k$ from
\reff{Xiphi}. Specifically, 
\bsplitl{ 
-\sum_{k,i,j=1}^d
	\frac{\partial}{\partial y_i}
	\brkts{
		{\rm m}_{ij}\brkts{
			\frac{\partial x_k}{\partial x_j}
			-\frac{\partial\xi^k_w}{\partial y_j}
		}\frac{\partial w_0}{\partial x_k}
	}
	& = \sum_{k,i,j=1}^d
	\frac{\partial}{\partial y_i}\brkts{
		{\rm m}_{ij}\brkts{
			\frac{\partial x_k}{\partial x_j}
			-\frac{\partial\xi^k_\phi}{\partial y_j}
		}\frac{\partial f(\phi_0)}{\partial x_k}
	}
\\&\quad
	- {\rm Pe}_{mic}\sum_{i=1}^d\brkts{{\rm u}^i - {\rm v}^i}\frac{\partial \phi_0}{\partial x_i}
	\quad\textrm{in $Y^1$}
	\,, 
}{XiwA} 
which can be simplified under a scale seperated 
chemical potential in the sense of Definition \ref{def:SSCP}, i.e.,
$\pd{}{x_k}f(\phi)
    = f'(\phi)\pd{\phi}{x_k}
    =\pd{w}{x_k}\
    \quad\textrm{for }1\leq k\leq d$,
to
the following cell problem,
\bsplitl{
\begin{cases}
-\sum_{i,j,k=1}^d
    \pd{}{y_i}\brkts{
        \delta_{ik}-\delta_{ij}\pd{\xi^k_w}{y_j}
    }f'(\phi_0)
\\\qquad\qquad
    =
    \lambda 
    \sum_{k,i,j=1}^d\pd{}{y_i}\brkts{
        {\rm m}_{ik}
        -{\rm m}_{ij}\pd{\xi^k_\phi}{y_j}
    }f'(\phi_0)
\\\qquad\qquad\quad
	- {\rm Pe}_{mic}\sum_{i=1}^d\brkts{{\rm u}^i - {\rm v}^i}
    &\textrm{ in }Y^1\,,
\\
\sum_{i,j,k=1}^d{\rm n}_i\Bigl(
        \brkts{
        \delta_{ij}\pd{\xi^k_w}{y_j}
        -\delta_{ik}
        }
\\\qquad\qquad
        -\lambda \sum_{k,i,j=1}^d\pd{}{y_i}\brkts{
            {\rm m}_{ik}
            -{\rm m}_{ij}\pd{\xi^k_\phi}{y_j}
        }
    = 0
    &\textrm{ on }\partial Y^1_w\cap\partial Y^2_w\,,
\\
\xi^k_w({\bf y})\textrm{ is $Y$-periodic and ${\cg M}_{Y^1}(\xi^k_w)=0$.}
\end{cases}
}{Xiw}

\medskip

A solvability constraint (e.g. the Fredholm alternative) immediately turns \reff{Xiw} 
into the following characterization of $\xi_w^k$ and ${\rm v}^i$, i.e.,
\bsplitl{
&{\rm v}^j
	:=\frac{\rm Pe_{mic}}{\av{Y^1}}\int_{Y^1}{\rm u}^j({\bf y})\,d{\bf y}
\\&
\begin{cases}
\quad
-\sum_{i,j,k=1}^d
    \pd{}{y_i}\brkts{
        \delta_{ik}-\delta_{ij}\pd{\xi^k_w}{y_j}
    }
    =
    \lambda 
    \sum_{k,i,j=1}^d\pd{}{y_i}\brkts{
        {\rm m}_{ik}
        -{\rm m}_{ij}\pd{\xi^k_\phi}{y_j}
    }
	&\textrm{ in }Y^1
\\\quad\quad
\sum_{i,j,k=1}^d{\rm n}_i\Bigl(
        \brkts{
        \delta_{ij}\pd{\xi^k_w}{y_j}
        -\delta_{ik}
        }
        -\lambda \sum_{k,i,j=1}^d\pd{}{y_i}\brkts{
            {\rm m}_{ik}
            -{\rm m}_{ij}\pd{\xi^k_\phi}{y_j}
        }
    = 0
    &\textrm{ on }\partial Y^1_w\cap\partial Y^2_w\,,
\\\quad
\xi^k_w({\bf y})\textrm{ is $Y$-periodic and ${\cg M}_{Y^1}(\xi^k_w)=0$.}
\end{cases}
}{Chrct}

We are then left to study the last problem \reff{O-0n} arising by the
asymptotic multiscale expansions. Problem \reff{O-0n}$_2$ for $\phi_2$ is
classical and leads immediately to the upscaled equation 
\bsplitl{
-\Delta_{\hat{\rm D}} \phi_0
    :=
    -{\rm div}\brkts{
        \hat{\rm D}\nabla \phi_0
    }
    = p w_0
    +\tilde{g}_0\,,
}{v_0}
see also \cite{Schmuck2012a}, where the porous media correction tensor $\hat{\rm D}:=\brcs{{\rm d}_{ik}}_{1\leq i,k\leq d}$
is defined by
\bsplitl{
\av{Y}{\rm d}_{ik}
	:= \sum_{j=1}^d\int_{Y^1}\brkts{
		\delta_{ik} - \delta_{ij}\frac{\partial\xi^k_\phi}{\partial y_j}
	}\,d{\bf y}\,.
}{dik}
Next, we apply the Fredholm alternative on equation \reff{O-0n}$_1$, i.e.,
\bsplitl{
\int_{Y^1}\Bigl\{
    \lambda\brkts{
         {\cg B}_2 w_0
        +{\cg B}_1 w_1
    }
    -\lambda{\cg B}_1\ebrkts{
        f(\phi_0)\phi_1
    }
    -\lambda{\cg B}_2f(\phi_0)
    -\partial_t{\cg A}_2^{-1}w_0
\\ \qquad\qquad
	-{\rm Pe}_{mic}\brkts{{\bf u}\cdot\nabla_{\bf y}}{\cg A}_2^{-1}w_2
	-{\rm Pe}_{mic}\brkts{({\bf u}-{\bf v})\cdot\nabla}{\cg A}_2^{-1}w_1
    \Bigr\}\,d{\bf y}
    =0\,.
}{SoCow2}
The term multiplied by $\lambda$ in \reff{SoCow2} can immediately
be rewritten by
\bsplitl{
\lambda\int_{Y^1}-\brkts{
		{\cg B}_2 w_0
        	+{\cg B}_1 w_1
	}\,d{\bf y}
    = -\lambda{\rm div}\brkts{
        \hat{\rm M}_w\nabla w_0
    }
    \,,
}{Bw}
where the effective tensor $\hat{\rm M}_w=\brcs{{\rm m}^w_{ik}}_{1\leq i,k\leq d}$ is defined by
\bsplitl{
{\rm m}^w_{ik}
    & :=
    \frac{1}{\av{Y}}\sum_{j=1}^d\int_{Y^1}\brkts{
        {\rm m}_{ik}
        -{\rm m}_{ij}\pd{\xi^k_w}{y_j}
    }\,d{\bf y}\,.
}{Mw}
The first term on the second line in \reff{SoCow2} transform
as in \cite{Schmuck2012a} to
\bsplitl{
-{\cg B}_1
	&\ebrkts{
		f'(\phi_0)\phi_1
    	}
\\&
	=	
	-\sum_{i,j=1}^d\ebrkts{
		\frac{\partial}{\partial x_i}\brkts{
			{\rm m}_{ij}f'(\phi_0)
			\sum_{k=1}^d\frac{\partial\xi^k_\phi}{\partial y_j}\frac{\partial\phi_0}{\partial x_k}
		}
		+\frac{\partial}{\partial y_i}\brkts{
			{\rm m}_{ij}f'(\phi_0)\sum_{k=1}^d\xi^k_\phi\frac{\partial^2\phi_0}{\partial x_k\partial x_j}
		}
	}\,,
}{Bphi}
where the last term in \reff{Bphi} disappears after integrating by parts. The first term 
on the right-hand side of \reff{Bphi} can be rewritten with the help of the chain rule
%
as follows
\bsplitl{
-{\cg B}_1
	&\ebrkts{
		f'(\phi_0)\phi_1
    	}
	=
	-\sum_{i,j=1}^d {\rm m}_{ij}\sum_{k=1}^d\frac{\partial\xi^k_\phi}{\partial y_j}
	\frac{\partial^2 f(\phi_0)}{\partial x_k\partial x_i}
\,.
}{B1}
After adding to \reff{B1} the term $-{\cal B}_2f(\phi_0)$, we can define a 
tensor $\hat{\rm M}_\phi=\brcs{{\rm m}^\phi_{ij}}_{1\leq i,k\leq d}$, i.e.,
\bsplitl{
{\rm m}^\phi_{ik}
    & :=
    \frac{1}{\av{Y}}\sum_{j=1}^d\int_{Y^1}\brkts{
	{\rm m}_{ik}
        -{\rm m}_{ij}\pd{\xi^k_\phi}{y_j}
    }\,d{\bf y}\,,
}{Mv}
such that
\bsplitl{
-{\cg B}_1
	\ebrkts{
		f'(\phi_0)\phi_1
    	}
	-{\cg B}_2f(\phi_0)
	= {\rm div}\brkts{
		\hat{\rm M}_\phi\nabla f(\phi_0)
	}
	\,.
}{B1B2}

The terms in the last line of \reff{SoCow2} become
\bsplitl{
-\frac{1}{\av{Y}}
	& \int_{Y^1}
		{\rm Pe}_{mic}\brkts{{\bf u}\cdot\nabla_{\bf y}}{\cg A}_2^{-1}w_2
		+{\rm Pe}_{mic}\brkts{({\bf u}-{\bf v})\cdot\nabla}{\cg A}_2^{-1}w_1
	\,d{\bf y}
\\
	&=
	-\frac{1}{\av{Y}}\int_{Y^1}
		{\rm Pe}_{mic}\brkts{{\bf u}\cdot\nabla_{\bf y}}\phi_2
	\,d{\bf y}
\\
	&\quad
	+\frac{{\rm Pe}_{mic}}{\av{Y}}\sum_{k,i=1}^d\int_{Y^1}({\rm u}^i-{\rm v}^i)
		\frac{\partial}{\partial x_i}\brkts{
			\delta_{ik}\xi^k_\phi({\bf y})\frac{\partial}{\partial x_k}\phi_0
		}
	\,d{\bf y}
	\,.
}{CnvTer}
Using the fact that ${\bf u}$ is divergence-free and after defining the tensor 
$\hat{\rm C}:=\brcs{{\rm c}_{ik}}_{1\leq i,k\leq d}$
by
\bsplitl{
{\rm c}_{ik}
	:= \frac{{\rm Pe}_{mic}}{\av{Y}}\int_{Y^1}({\rm u}^i-{\rm v}^i)
			\delta_{ik}\xi^k_\phi({\bf y})
	\,d{\bf y}\,,
	&
}{Tnsrs}
we finally obtain with the previous considerations the following upscaled phase field equation
\bsplitl{
p\pd{{\cg A}_2^{-1}w_0}{t}
   &  = {\rm div}\Bigl(
        \Bigl[
	\lambda\hat{\rm M}_\phi f'(\phi_0)
	+\hat{\rm C}
        \Bigr]\nabla \phi_0
	\Bigr)
\\&\quad
    -\frac{\lambda}{p}{\rm div}\brkts{
        \hat{\rm M}_w\nabla \brkts{
            {\rm div}\brkts{
                \hat{\rm D}\nabla \phi_0
            }
            -\tilde{g}_0
        }
    }
\,.
}{EfW0}
Using \reff{v_0} finally leads to the effective macroscopic phase field equation 
\bsplitl{
p\pd{\phi_0}{t}
   &  = {\rm div}\Bigl(
        \Bigl[
	\lambda\hat{\rm M}_\phi f'(\phi_0)
	+\hat{\rm C}
        \Bigr]\nabla \phi_0
	\Bigr)
\\&\quad
    -\frac{\lambda}{p}{\rm div}\brkts{
        \hat{\rm M}_w\nabla \brkts{
            {\rm div}\brkts{
                \hat{\rm D}\nabla \phi_0
            }
            -\tilde{g}_0
        }
    }
	\,.
}{EfMaPhFi}

\csection{Proof of Theorem \ref{thm:2}}\label{sec:Thm2}
The proof follows in three basic steps: In Step 1, we establish a priori estimates that provide 
compactness required for Step 2 where we construct a sequence of approximate solutions. In Step 3, we pass to the 
limit in the sequence of approximate solutions which provide then existence and uniqueness of the original problem.
\\
\emph{Step 1: (A priori estimates)} i) Basic energy estimate: As we establish the existence and uniqueness of weak solutions, we 
rewrite \reff{UpScPrTh} in the sense of distributions, i.e., for all 
$\varphi\in H^2_E$ it holds that
\bsplitl{
\frac{p}{2}\frac{\partial}{\partial t}\brkts{
		\phi_0,\varphi
	}
	& +\frac{\lambda}{p}\brkts{
		{\rm div}\brkts{
			\hat{\rm D}\nabla\phi_0
		},
		{\rm div}\brkts{
			\hat{\rm M}_w\nabla\varphi
		}
	}
	+\frac{\lambda}{p}\brkts{
		\hat{\rm M}_w\nabla\tilde{g}_0,\nabla\varphi
	}
\\&
	+\brkts{
		\lambda f'(\phi_0)\hat{\rm M}_\phi\nabla\phi_0,
		\nabla\varphi
	}
	+ \brkts{
		\hat{\rm C}\nabla\phi_0,
		\nabla\varphi
	}
	= 0
	\,.
}{WFupPhFi}
Note that for an isotropic mobility of the form $\hat{\rm M}=m\hat{\rm I}$, where $\hat{\rm I}$ is the identity 
tensor, the following identity holds $m\hat{\rm D}=\hat{\rm M}_w$. For simplicity, we will base our proof on 
this identity. The general case is then verified along the same lines using properties of symmetric positive definite 
tensors. Using the notation $\Delta_{\hat{\rm D}}:={\rm div}\brkts{\hat{\rm D}\nabla}$ and the test 
function $\varphi=\phi_0$ leads to the following inequality
\bsplitl{
\frac{p}{2}\frac{d}{dt}\Ll{\phi_0}^2
	& +\frac{\lambda m}{p}\Ll{\Delta_{\hat{\rm D}}\phi_0}^2
	+\lambda b_{2r}m_\phi\int_\Omega\phi_0^{2r-2}\av{\nabla\phi_0}^2\,d{\bf x}
\\
	& \leq C\Ll{\nabla\phi_0}^2
		+\frac{2\lambda m_w}{p}\Ll{\nabla\tilde{g}_0}\Ll{\nabla\phi_0}
	\leq C\Ll{\nabla\phi_0}^2
		+\frac{\lambda}{2p}m_w\Ll{\nabla\tilde{g}_0}^2
	\,,
}{ApEs1}
where we applied the existence of a constant $C>0$ such that 
$f(s)s\geq pb_{2r}s^{2r}-C\,,$ for all $s\in\mathbb{R}$ due to 
Assumption (PF). With the equivalence of norms in $H^2(\Omega)$, i.e., 
$\N{\phi_0}{H^2(\Omega)}
	\leq C\brkts{
		\Ll{\Delta \phi_0}+\Ll{m(\phi_0)}
	}$\,,
and interpolation estimates we obtain
\bsplitl{
\Ll{\nabla\phi_0}^2
\	& \leq C\Ll{\phi_0}\N{\phi_0}{H^2(\Omega)}
	\leq C\Ll{\phi_0}\brkts{
		\Ll{\Delta\phi_0}+\alpha
	}
	\leq C\Ll{\phi_0}\brkts{
		\Ll{\Delta_{\hat{\rm D}}\phi_0}+\alpha
	}
\\&
	\leq \frac{\kappa}{2}\Ll{\Delta_{\hat{\rm D}}\phi_0}^2
		+\frac{1}{C\kappa}\Ll{\phi_0}^2
		+\frac{\kappa\alpha^2}{2}
	\,,
}{auxEs}
where $m(\phi_0):=\frac{1}{\av{\Omega}}\int_\Omega\phi_0\,d{\bf x}$ with 
$m(\phi_0)\leq\alpha$. This 
leads to 
\bsplitl{
\frac{p}{2}\frac{d}{dt}\Ll{\phi_0}^2
	&
	+ \brkts{\frac{\lambda m}{p}-\frac{\kappa}{2}}\Ll{\Delta\phi_0}^2
	+ \lambda b_{2r}\int_\Omega \phi^{2r-2}\av{\nabla\phi_0}^2\,d{\bf x}
\\&
	\leq
	C(\kappa)\Ll{\phi_0}^2
	+C(\alpha,\kappa)
	+\frac{\lambda}{2p}m_w\Ll{\nabla\tilde{g}_0}^2
	\,,
}{Es1}
which turns with Gronwall's inequality into the expression
\bsplitl{
\Ll{\phi_0(\cdot,t)}^2
	\leq \Ll{\phi_0(\cdot,0)}^2{\rm exp}\brkts{
		Ct
	}
	+\int_0^t\brkts{C(\alpha)+\frac{\lambda}{2p}m_w\Ll{\nabla\tilde{g}_0}^2}
		{\rm exp}\brkts{-Cs}
	\,ds
\,,
}{Es2}
for $T^*\geq 0$ 
such that we finally obtain $\phi\in L^\infty(0,T^*;L^2(\Omega))\cap L^2(0,T^*;H^2_E(\Omega))$.
\\
ii) Control over time derivative: Using the test function $\varphi\in H^2_E(\Omega)$ in \reff{WFupPhFi} 
allows us to estimate the time derivative term by
\bsplitl{
\frac{p}{2}\av{\brkts{
		\partial_t\phi_0,\varphi
	}}
	&
	\leq C\brkts{
		\Ll{\Delta_{\hat{\rm D}}\phi_0}
		+\Ll{\tilde{g}_0}
		+\Ll{f(\phi_0)}
		+\Ll{\phi_0}
	}\Ll{\Delta\varphi}
	\,,
}{TDest}
such that 
$\N{\partial_t\phi_0}{(H^2_E(\Omega))^*}=\sup_{\varphi\in H^2_E}\frac{\av{\brkts{\partial_t\phi_0,\varphi}}}{\N{\varphi}{H^2_E}}\leq C$
and hence $\N{\partial_t\phi_0}{(H^2_E(\Omega))^*}^2\leq C(T^*)$.
\\
\emph{Step 2: (Galerkin approximation)} As $H^2_E(\Omega)$ is a separable Hilbert space, we can 
identify a linearly independent basis $\varphi_j\in H^2_E(\Omega)$, $j\in\mathbb{N}$, 
which is complete in $H^2_E(\Omega)$. This allows us to define for each $N\in\mathbb{N}$ 
approximate solutions $\phi_0^N=\sum_{j=1}^N\eta^N_j(t)\varphi_j$ which solve 
\bsplitl{
\frac{p}{2}\frac{\partial}{\partial t}\brkts{
		\phi_0^N,\varphi_j
	}
	& +\frac{\lambda m}{p}\brkts{
		{\rm div}\brkts{
			\hat{\rm D}\nabla\phi_0^N
		},
		{\rm div}\brkts{
			\hat{\rm D}\nabla\varphi_j
		}
	}
	+\frac{\lambda}{p}\brkts{
		\hat{\rm M}_w\nabla\tilde{g}_0,\nabla\varphi_j
	}
\\&
	+\brkts{
		\lambda f'(\phi_0^N)\hat{\rm M}_\phi\nabla\phi_0^N,
		\nabla\varphi_j
	}
	+ \brkts{
		\hat{\rm C}\nabla\phi_0^N,
		\nabla\varphi_j
	}
	= 0
	\quad
	j=1,\dots\,,N
	\,,
}{ApSol}
for the initial condition $\phi_0^N(\cdot,0)=\psi^N(\cdot)$. The initial value problem \reff{ApSol} 
represents a system of $N$ ordinary differential equations (ODEs) for the coefficients $\eta_j^N(t)$. Hence, 
classical ODE theory immediately provides existence and uniqueness of $\phi^N_0$. Moreover, we have 
$\phi_0^N\in C(0,T^*;H^2_E(\Omega))$ and $\partial_t\phi_0^N\in L^2(0,T^*;H^{-2}_E(\Omega))$.
\\
\emph{Step 3: (Passing to the limit)} In the same way as in \emph{Step 1}, we can 
derive a priori estimates for the approximate solutions $\phi_0^N$ by using the test 
function $\phi_0^N$ instead of the basis functions $\varphi_j$ in \reff{ApSol}. Hence, by weak 
compactness, see Theorem \ref{thm:AL} in Section \ref{sec:Prelim}, there exist subsequences 
(for simplicity still denoted by $\phi_0^N$) such that 
\bsplitl{
& \phi_0^N \rightharpoonup \phi_0
	\quad\textrm{in }L^2(0,T^*;H^2_E(\Omega))
	\textrm{ weakly}
	\,,
\\&
\phi_0^N \stackrel{*}{\rightharpoonup} \phi_0
	\quad\textrm{in }L^\infty(0,T^*;L^2(\Omega))
	\textrm{ weak-star}
	\,,
\\&
\phi_0^N \to \phi_0
	\quad\textrm{in }L^2(0,T^*;L^2(\Omega))
	\textrm{ strongly}
	\,,
}{WCV}
for $N\to\infty$.
This allows us to pass to the limit in the initial value problem \reff{ApSol} such 
that we obtain 
\bsplitl{
\frac{p}{2}\frac{\partial}{\partial t}\brkts{
		\phi_0,\varphi
	}
	& +\frac{\lambda}{p}\brkts{
		{\rm div}\brkts{
			\hat{\rm D}\nabla\phi_0
		},
		{\rm div}\brkts{
			m\hat{\rm D}\nabla\varphi
		}
	}
	+\frac{\lambda}{p}\brkts{
		\hat{\rm M}_w\nabla\tilde{g}_0,\nabla\varphi
	}
\\&
	+\brkts{
		\lambda f'(\phi_0)\hat{\rm M}_\phi\nabla\phi_0,
		\nabla\varphi
	}
	+ \brkts{
		\hat{\rm C}\nabla\phi_0,
		\nabla\varphi
	}
	= 0
	\,,
}{LP}
for all $\varphi\in H^2_E(\Omega)$. In the same way we can pass to the limit 
with respect to the initial condition $\psi^N$.
\hfill$\Box$

\csection{Conclusions}\label{sec:Cncls}
The main new result here is the extension of the the study by Schmuck
\emph{et al.} in the absence of flow~\cite{Schmuck2012a} to include a
periodic fluid flow in the case of sufficiently large P\'eclet number. The
resulting new effective porous media approximation \reff{EffImFlThm} of the
microscopic Stokes-Cahn-Hilliard problem \reff{PeFl}--\reff{MiPr} reveals
interesting physical characteristics such as diffusion-dispersion relations
by \reff{kappaThm}$_2$--\reff{kappaThm}$_3$ for instance. The homogenization
methodology developed here allows for the systematic and rigorous derivation
of effective macroscopic porous media equations as motivated for instance in \cite{Hornung1997} 
where the author identifies Darcy's law after upscaling the Stokes equations. By a slight modification 
of the homogenization method as initiated in \cite{Allaire2010a,Brenner1980,Marusic-Paloka2005} 
for diffusion-convection problems, one recovers here  for the first time, to the best knowledge of the authors, rigorous and systematic diffusion-dispersion relations for general Stokes-Cahn-Hilliard equations.

We note that the Cahn-Hilliard and related equations generally model more
complex material transport \cite{NovickCohen2008,Promislow2009} than
classical Fickian diffusion. In this context, our result of an upscaled
convective Cahn-Hilliard equation hence provides diffusion-dispersion
relations for generalized non-Fickian material transport, that is, not just
the product of a gradient of particle concentration and a constant diffusion
matrix.

Our current periodicity assumption on the fluid flow imposed by \reff{PeFl} 
implies a quasi-steady
state on the fluid velocity and seems inevitable for the
homogenization theory to work such as the assumption of a scale separated
chemical potential on the level of the reference cells.

So far, we are restricted to mixtures of two fluids by our model. An
extension towards mixtures of $N>2$ components is studied in \cite{Otto1997}
where the authors compare a local and non-local model for incompressible
fluids. In specific applications, it might be of interest to extend the here
developed framework towards such multi-component mixtures.

Finally, the rigorous and systematic derivation of effective macroscopic
immiscible flow equations provides a promising and convenient alternative in
view of the broad applicability of the Cahn-Hilliard equations. The strength
of our approach is based on its foundation on a thermodynamically motivated
homogeneous free energy which is generally derived on systematic physical
grounds. Moreover, even if a systematic derivation is impossible, one can
mathematically design such a free energy based on physical principles and
phenomenological observations. As an example of the applicability
of the presented upscaled equations, we have numerically solved the
homogenised Cahn-Hilliard equation for a porous medium defined by 
periodic reference cells consisting of a characteristic perturbed straight channel. We
observe that the macroscopic solution retained the geometric characteristics
induced at the microscopic scale. This numerical study represents a first
step towards the use of the presented methodology in more complex geometries
with e.g.~non-periodic properties, something that we leave as a future work.
In addition, a more detailed numerical study shall allow to rationally
analyze current applications in science and engineering and hopefully reveal
potential new ones.

\section{Acknowledgements}
We acknowledge financial support from EPSRC Grant No.
EP/H034587, EPSRC Grant No. EP/J009636/1, EU-FP7 ITN Multiflow and ERC Advanced Grant No. 247031.

\bibliographystyle{plainnat}
\bibliography{effPhaseFieldFlow_main5}

          %

\end{document}